\documentclass[
    reprint, amsmath, amssymb, aps, pra
]{revtex4-2}

\usepackage{graphicx}
\usepackage{dcolumn}
\usepackage{bm}
\usepackage{soul}

\usepackage{xcolor}

\newcommand{\braket}[2]{\langle #1 | #2 \rangle}
\newcommand{\brakket}[3]{\langle #1 | #2| #3 \rangle}

\newcommand{\ket}[1]{| #1 \rangle}

\renewcommand{\d}{{\rm d}}

\newcommand{\Tr}{{\rm Tr}}
\newcommand{\pd}{\partial}
\newcommand{\mc}{\mathcal}
\newcommand{\bnabla}{\bm{\nabla}}

\newcommand{\revision}[1]{{\color{olive}#1}}
\renewcommand{\revision}[1]{#1}

\begin{document}

\title{Negative thermodynamic pressure: no-go theorem and yes-go examples}

\author{Varazdat Stepanyan}
\affiliation{Physics Institute, Yerevan State University, Yerevan, Armenia}

\begin{abstract}
    Theory and experiment have long discussed negative thermodynamic pressure states, but their microscopic origins are unclear. I address this problem within the framework of quantum thermodynamics. I show that the pressure exerted on the boundary is positive when there is no interaction with the boundary. This is formalized by a no-go theorem that holds for any quantum state \revision{with finite motion. As a consequence of this analysis, I deduce a general formula for quantum non-equilibrium pressure.} It is believed that stable negative pressure states cannot exist in gases. I provide solvable examples of quantum and classical gases, where negative pressure is achieved due to a suitable coupling with the boundary walls.
\end{abstract}

\maketitle

\section{Introduction}

    The existence and physical meaning of thermodynamic negative pressure states have long been subjects of debate. Negative pressure states were discussed theoretically \cite{landau5,imre98} with opposing claims on the possibility of the existence of such states. Laboratory experiments suggest the existence of metastable states with negative pressures in liquids \cite{balibar00,caupin13,duffner18, stiller23}. In plant and animal physiology, negative pressure states were also proposed as effective mechanisms for fluid transport \cite{brown13, miller68}. Such mechanisms are employed in artificial trees and similar devices \cite{hongya20}. In field theory, negative pressure relates to quantum vacuum \cite{casimir48, trunov90, milton01, stabile17}. It is still unclear how stable negative pressure states can exist in quantum statistical mechanics. Negative pressure and negative temperature are both extreme conditions in thermodynamics, but the latter (compared to the former) has been extensively studied both theoretically and experimentally \cite{onsager,purcell,ramsey56,berdi,hsu,braun,hilbert14,warren15,penrose17}.
    
    Pressure determines the amount of work done during a slow, thermally isolated (adiabatic) or isothermal process \cite{landau5}. The volume here is defined via infinite potential walls. These walls can be the separation layer between the system and its environment, the walls of a cavitation (bubble), {\it etc}. 
    
    To solve the pressure sign problem, first I recite the known thermodynamic argument for positive pressure and then derive this result from equilibrium classical statistical mechanics. I point out two main limitations of such derivations: they do not account for particle-wall interactions and do not allow for non-equilibrium. Then I continue with quantum mechanics and deduce a general equation for pressure in any non-equilibrium state and for any Hamiltonian. I state and prove a no-go theorem for quantum systems where stable negative pressures are impossible. An important consequence of the theorem is that negative pressure states require specific interactions between the system and the walls (other than their confining feature). 
    I discuss two (quantum and classical) models of an ideal gas with a particle-wall interaction and show that negative pressure states are possible, despite a wide belief that gases do not show negative pressure \cite{imre98, imre07, duffner18}. 
    Finally, I point out that the theorem contradicts conclusions drawn about the negative pressure from the van der Waals phenomenological equation of state. I suggest a way of removing this contradiction that will enlarge our understanding of the van der Waals model.

\section{Thermodynamic analysis}
    One of the standard arguments against the existence of stable negative pressure states is as follows. Start with the first law $\d E=T\d S-P\d V$ for a macroscopic equilibrium system \cite{landau5}. Notations here are standard: energy, temperature, entropy, thermodynamic pressure, volume. We always assume $T>0$. The first law implies for the pressure: 
    \begin{align}
    P=-(\partial E/\partial V)|_S=T(\partial S/\partial V)|_E.
        \label{paper-as}
    \end{align}
    For spontaneous processes in the isolated situation ($\d E=0$) we use the last formula in \eqref{paper-as}. Now for such processes, we have $\partial S>0$ due to the second law. Let us now assume that $\partial V>0$ in \eqref{paper-as}. Recall that $T>0$. Now $\partial S>0$ and $\partial V>0$ in \eqref{paper-as} imply $P>0$ \cite{landau5}. Such a state can be stable if restricted by external walls. If $\partial V<0$ (and $T>0$), the system \revision{is unstable and spontaneously decreases the volume reaching a self-sustained state, where it} does not explode or collapse. This means that \revision{it reaches zero pressure}. Overall, this shows $P\geq 0$ \cite{landau5}. 
    
    Now $P\geq 0$ can also be derived for concrete ensembles. The equilibrium state of a closed, ergodic classical system with Hamiltonian $\mc{H}(\Gamma)$ is described by a microcanonical probability density $\rho_{\rm M}(\Gamma)\propto \delta (E-\mc{H}(\Gamma))$, where $\Gamma$ is the phase-space point and $E$ is the fixed energy. In a thermally isolated slow (adiabatic) process, the entropy ${S} = \ln{\int \Theta(E - \mc{H}(\Gamma))\d\Gamma}$ stays constant, where $\Theta(x)$ is the Heaviside step function \cite{hertz1910,kasuga1961,saakian07,warren15}. Now apply the second relation in \eqref{paper-as} to two microcanonic states having the same energy $E$, but different volumes $V$ and $V+\d V$ with $\d V>0$. We assume that $\mc{H}(\Gamma)$ does not depend on the coordinates of the walls. Hence all phase-space points that contributed to $S(V)$ will also contribute to $S(V+\d V)$. We conclude that $S(V+\d V)\geq S(V)$, and thus $P\geq 0$ from \eqref{paper-as}. \revision{The same result can be achieved using the Boltzmann entropy $S=\ln\int\delta(E-\mc{H}(\Gamma))\d\Gamma$ instead of Gibbs entropy. I will refrain from delving into the discussion of the differences between the two entropies as it is not relevant to the points of this paper.}
    
    For a canonic state at a constant temperature $T$, $\rho_{\rm C}(\Gamma)=\frac{1}{Z} e^{-\mc{H}(\Gamma)/T}$, we employ the first law and the definition $F=-T\ln Z$ of the free energy for deducing:
    \begin{align}
    P=-\frac{\partial F}{\partial V}\Big|_T=T\frac{\partial \ln Z}{\partial V}
    =T\frac{\partial }{\partial V}\ln \int e^{-\mc{H}(\Gamma)/T}
    \d\Gamma,
        \label{paper-free}
    \end{align}
    where $Z$ is the statistical sum. Assuming that $\mc{H}(\Gamma)$ does not depend on the coordinates of the walls, we get that $Z$ is a growing function of $V$. Hence, $P\geq 0$ from \eqref{paper-free}.
    
    In each of the three derivations above, there are two limitations. {\it (i)} They assume that the walls do not interact with the system in any way other than confining it. Such interactions can lead to a state where reducing $V$ (keeping the wall fixed) will require work, but moving the walls to reduce $V$ will extract work from the system. This can produce a stable \revision{(i.e., finite-motion)} absolute negative pressure state, as shown below. {\it (ii)} The derivations apply to equilibrium states only. Since metastable states are not in equilibrium, limitation {\it (ii)} led to an opinion (widely echoed, but unwarranted, as shown below) that metastable states can be prone to a negative pressure \cite{landau5}.

\section{General formula for the quantum non-equilibrium pressure}
    Consider a $N$-particle quantum system inside an infinite potential well confining the system in the convex coordinate set $\Omega$ (support) with volume $V$. The Hamiltonian of this system is
    \begin{equation}\label{paper-eq:hamiltonian}
        \mc{H} = -\frac{1}{2}\Delta + \mc{U}(\bm{x};\Omega),
    \end{equation}
    where $m=\hbar=1$, $\bm{x}_j$ is the $D$-dimensional coordinate of $j$-th particle, $\bm{x}=(\bm{x}_1...\bm{x}_N)$ is an $ND$ dimensional vector, and $\bnabla$, $\Delta=\sum_{j=1}^{N} (\partial/\partial \bm{x}_j)^2$ are the gradient and Laplace operators of that $ND$ dimensional space. $\mc{U}$ includes inter-particle interactions $\mc{U}_{\rm in}$, external potentials, and the potential ${U}_{\rm ex}$ created by the confining walls:
    \begin{equation}
        \mc{U}(\bm{x}_1...\bm{x}_N;\Omega) = \mc{U}_{\rm in}(\bm{x}_1...\bm{x}_N)+{\sum}_{j=1}^nU_{\rm ex}(\bm{x}_j;\Omega),
    \label{paper-borneo}
    \end{equation}
    where ${U}_{\rm ex}$ depends on $\Omega$, but does not include the infinite potential of the walls. That term is accounted for by the Dirichlet boundary conditions, which holds for all states including energy eigenstates ($\pd\Omega$ is the boundary of $\Omega$): 
    \begin{equation}
    \label{paper-bou}
    \mc{H}\ket{n}=E_n\ket{n},\quad 
    \braket{n}{\bm x \in \pd\Omega}=0.
    \end{equation}
    The system is described by a density matrix $\rho(t)$. 
    \revision{Eq. \eqref{paper-bou} and the density matrix $\rho(t)$ refer to the finite motion, where $\rho(t)$ is integrable for large and small distances. Such a system neither collapses nor spreads into infinity.} 
    
The pressure is defined by the amount of work extracted from the system when increasing the volume $V$ of the system in a thermally isolated process. Thus, the support set moves from $\Omega$ at the initial time $t_{\rm i}$ to $\Omega'$ at the final time $t_{\rm f}$. 
    \revision{As for any thermally isolated process}, the work $W$ is equal to the difference of the average energy \cite{allahverdyan05}:
    \begin{align}
    & W=E(t_{\rm f})-E(t_{\rm i}),\quad
    E(t_{\rm i})=\Tr{(\rho(t_{\rm i})\mc{H})}={\sum}_n p_n E_n, \nonumber\\ 
    &\brakket{n}{\rho(t_{\rm i})}{n}\equiv p_n, \quad 
    E(t_{\rm f})=\Tr{(\rho(t_{\rm f})\mc{H}(\Omega'))}.
    \label{paper-energy}
    \end{align}
    For defining the pressure $P$ we additionally assume that the process is slow (adiabatic). Such processes are work-optimal in quantum thermodynamics \cite{allahverdyan05}. Thus, we get
    \begin{equation}\label{paper-def:press}
        P = -(\pd E/\pd V)|_{\rm ad}.
    \end{equation}
    We will calculate \eqref{paper-def:press} under three further assumptions. 

    {\bf 1.} $\Omega$ and $\Omega'$ are convex sets. $\Omega'$ embeds $\Omega$.
    
    {\bf 2.} $\Omega'$ is obtained from $\Omega$ via a scaling relation $\bm{x}_j\to (1+\alpha) \bm{x}_j$, where $\alpha>0$ is small \cite{zubarev1973}. 

    \begin{figure}[ht]
        \centering
        \includegraphics[width=0.45\textwidth]{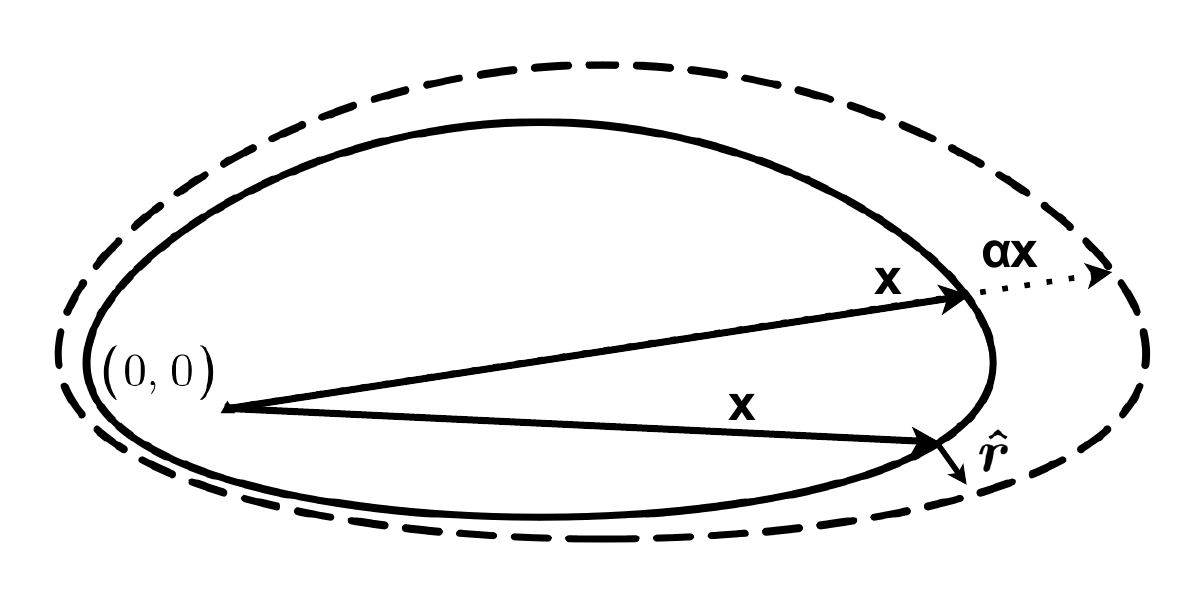}
        \caption{\revision{A schematic diagram of the sets $\Omega$ (solid) and $\Omega'$ (dotted) as well as the outwards normal $\hat{r}$ used in (\ref{paper-eq:interm2}, \ref{paper-eq:press_quant}).}}
        \label{paper-fig:convex}
    \end{figure}
    
    I believe {\bf 1} is essential, while {\bf 2} is technical and shall be generalized in the future. Once the coordinate origin is arbitrary, the class of $\Omega'$'s allowed by {\bf 2} is still large. 
    
    {\bf 3.} The dynamics of $\rho$ in this adiabatic process holds the quantum adiabatic theorem \cite{migdal,oh07}: 
    \begin{align}
    \label{paper-ado}
    p_n=\langle n,\Omega |\rho(t_{\rm i})|n,\Omega\rangle=\langle n,\Omega' |\rho(t_{\rm f})|n,\Omega'\rangle, 
    \end{align}
    where $|n,\Omega'\rangle$ is the eigenvalue of $\mc{H}(\Omega')$ with $\braket{n,\Omega'}{\bm x \in \pd\Omega'}=0$. \revision{The standard formulation of the adiabatic theorem requires that energy levels with widely different populations do not approach each other (avoided crossing). Otherwise, the known statement of the theorem is destroyed by Landau-Zener transitions \cite{migdal,oh07}. 
    At this point, one may conclude that the adiabatic theorem does not apply to macroscopic systems with densely distributed energy levels. Such a conclusion would be premature, because the basic statement of the adiabatic theorem is valid as well for ergodic (as well as integrable) classical systems performing finite motion \cite{hertz1910,kasuga1961,ott1979goodness, saakian07}. Note that the assumption of finiteness (i.e., stability) is important, since even the classical adiabatic theorem is violated to various degrees, if this assumption is not met \cite{allahverdyan07}. With such reservations, I will continue using the quantum adiabatic theorem \eqref{paper-ado}.}
        
    We get from (\ref{paper-energy}, \ref{paper-def:press}, \ref{paper-ado}):
    \begin{align}
                P =& -\frac{1}{DV}\lim_{\alpha\to 0}{\sum}_{n}\, p_n \, \frac{1}{\alpha}\times\nonumber \\
                &\Big(\brakket{n,\Omega'}{\mc{H}(\Omega')}{n,\Omega'}-\brakket{n,\Omega}{\mc{H}(\Omega)}{n,\Omega} \Big).
                \label{paper-kro}
    \end{align}
    Eq. \eqref{paper-kro} is also valid for \revision{an open system described by a Gibbs density in a slow isothermal process; cf. \eqref{paper-free}. For this particular case, assumption \textbf{3} is not necessary, and we only need the first two assumptions.} Now $\rho$ is a thermal (Gibbs) state, and the pressure is defined via the free energy; see \eqref{paper-free}. This agreement between isothermal and adiabatic processes is due to the employed first order of perturbation theory; see  \S 1 of supplementary material (S.M.) \cite{supp}. \revision{This agreement also implies that the equilibrium pressure is a state function and does not depend on the process.}
    
    Recalling {\bf 2}, we change in \revision{the left term of} \eqref{paper-kro} \revision{the} coordinates as $\bm{x} \to \frac{\bm{x}}{1+\alpha}$ and arrive back to the original integration over $\Omega$ with boundary conditions \eqref{paper-bou} \cite{migdal}. Thereby we obtain a perturbation for Hamiltonian \eqref{paper-eq:hamiltonian}
    \begin{align}
        \delta\mc{H} =& \alpha\Delta+\alpha\,\bm{x\cdot\nabla} \mc{U} \nonumber\\
        +& {\sum}_{j=1}^N\big[U_{\rm ex}(\bm{x}_j;\Omega')-U_{\rm ex}(\bm{x}_j;\Omega)\big] + \mc{O}(\alpha^2).
        \label{paper-mali}
    \end{align}
    Eqs. (\ref{paper-eq:hamiltonian}, \ref{paper-borneo}, \ref{paper-kro}, \ref{paper-mali}) imply for the pressure
    \begin{align}
        &P = -\frac{1}{DV}{\sum}_np_n\bigg[\bigg(\brakket{n}{\Delta}{n} + \brakket{n}{\bm{x}\bnabla\mc{U}}{n}\bigg) - P_{\rm ex}^{(n)}\bigg],\nonumber\\
        &P_{\rm ex}^{(n)}= {\sum}_{j=1}^N\lim_{\alpha\to0}\frac{\brakket{n}{U_{\rm ex}(\bm{x}_j;\Omega)-U_{\rm ex}(\bm{x}_j;\Omega')}{n}}{\alpha},
        \label{paper-dag}
    \end{align}
    where the expression inside round brackets is a modified version of the virial theorem \cite{terletskii79}. \revision{To simplify it first we integrate by parts
    \begin{equation}\begin{split}
        &-\int_\Omega \bm{x} \bnabla \mc{U} \psi^2_n d\bm{x} = ND \brakket{n}{\mc{U}}{n} -\int_\Omega \bm{x} \mc{U} \bnabla\psi_n^2\d\bm{x} =\\
        &NDE_n+\frac{ND}{2}\brakket{n}{\Delta}{n}+\int_\Omega\Delta\psi\bm{x}\bnabla\psi\d\bm{x}+E_n \int_\Omega \bm{x}\bnabla\psi_n^2,
    \end{split}\end{equation}
    where $E_n \psi_n + \frac{1}{2}\Delta \psi_n = \mc{U}\psi_n$ and $\psi_n(\bm{x}) = \braket{\bm{x}}{n} = \braket{n}{\bm{x}}$ is real because the Hamiltonian itself is real. Once again integrating by parts the last term we see that $E_n\int_\Omega \bm{x}\bnabla\psi_n^2 = -NDE_n$ and, therefore, the term in round brackets in equation \eqref{paper-dag} becomes
    \begin{equation}\begin{gathered}\label{paper-eq:interm}
        -\brakket{n}{\Delta}{n} - \brakket{n}{\bm{x}\bnabla\mc{U}}{n} = \\
        =\frac{ND-2}{2}\brakket{n}{\Delta}{n}+\int_{\Omega} \bm{x}\bnabla\psi_n\Delta\psi_n \d\bm{x},
    \end{gathered}\end{equation}
    }
    Eq. \eqref{paper-eq:interm} is \revision{also} valid for a static external electromagnetic field \revision{now with the term $\frac{1}{2}\int_\Omega\bm{x}\big(\bnabla\psi_n^*\Delta\psi_n + \bnabla\psi_n\Delta\psi_n^*\big)\d\bm{x}$}. The terms with electromagnetic fields cancel out and we arrive to the same result \revision{with a complex wave function}; see S.M. \S 2 \cite{supp} where \eqref{paper-eq:press_quant} is derived for an external static electromagnetic field. 
    We integrate the second term of \eqref{paper-eq:interm} by parts and get 3 terms: \revision{
    \begin{equation}\begin{gathered}\label{paper-eq:interm2}
        \int_{\Omega} \bm{x}\bnabla\psi_n\Delta\psi_n \d\bm{x} = \int_{\pd\Omega} (\bm{x}\bnabla\psi_n)(\bm{\hat{r}}\bnabla\psi_n)\d s -\\
        -\int_\Omega \bnabla(\bm{x}\bnabla\psi_n)\bnabla\psi_n\d\bm{x} = \int_{\pd\Omega} (\bm{x}\bm{\hat{r}})|\bnabla\psi_n|^2\d s +\\
        + \brakket{n}{\Delta}{n} - \frac{1}{2}\int_\Omega \bm{x}\bnabla|\bnabla\psi_n|^2\d\bm{x},
    \end{gathered}\end{equation}
    where $\bm{\hat{r}}$ is the outwards normal vector of $\pd\Omega$.} Here employ that $\psi_n(\bm{x}\in\pd\Omega)=0$ is constant over $\partial\Omega$, which means 
    $\bnabla\psi_n|_{\partial \Omega}=\bm{\hat{r}}(\bnabla\psi_n\bm{\hat{r}})|_{\partial \Omega}$ and $(\bm{\hat{r}}\bnabla\psi_n)^2|_{\partial \Omega}=(\bnabla\psi_n)^2|_{\partial \Omega}$. \revision{The last term in \eqref{paper-eq:interm2} can be further simplified by integrating by parts again, getting $-\frac{1}{2}\int_{\pd\Omega}(\bm{x}\bm{\hat{r}})|\bnabla\psi_n|^2\d s - \frac{ND}{2}\brakket{n}{\Delta}{n}$. Finally, grouping similar terms this chain of integration by parts results in
    \begin{equation}
        -\brakket{n}{\Delta}{n} - \brakket{n}{\bm{x}\bnabla\mc{U}}{n} = \frac{1}{2}\int_{\pd\Omega}(\bm{x}\bm{\hat{r}})|\bnabla\psi_n|^2\d s
    \end{equation}
    and we get a general equation for pressure}.
    \begin{equation}\label{paper-eq:press_quant}
        P = \frac{1}{DV}{\sum}_np_n\bigg[\frac{1}{2}\oint_{\pd\Omega} (\bm{x}\bm{\hat{r}})|\bnabla\psi_n|^2\d s + P_{\rm ex}^{(n)}\bigg].
    \end{equation}
    Since the origin of our coordinates lies within the convex set $\Omega$, for a point $\bm x$ on the boundary $\pd\Omega$ we have $L(\bm{x})\equiv\bm{x}\bm{\hat{r}}\geq0$, as can be verified graphically \revision{Fig. \ref{paper-fig:convex}}. Therefore, the first term of the equation \eqref{paper-eq:press_quant} is positive. A similar result without $P_{\rm ex}^{(n)}$ has been derived in a different way for a single one-dimensional particle inside square walls without interactions with the walls \cite{francisco23}. Positive pressure has also been shown for a many-body system at zero temperature with delta function interactions \cite{llano83}, showing that the lowest energy eigenvalue is decreasing.
        
    \subsection{No-go theorem}
        \textit{A quantum system without interaction with the walls confining it within a volume cannot have stable negative pressure states \revision{Here stability is understood in the sense of a finite motion; cf. the discussion after \eqref{paper-bou}.}}
                
        The interaction with the walls in our derivation was governed by the potential $U_{\rm ex}$. Removing it in \eqref{paper-eq:press_quant} makes $P_{\rm ex}^{(n)}=0$ and we get $P\geq 0$ from \eqref{paper-eq:press_quant}, as argued above. The theorem implies that a stable system always fills the volume confined by the walls never collapsing to a smaller volume. Another consequence is that the energy eigenvalues are non-increasing functions of the coordinates of the boundary $\pd\Omega$. A similar result has been proven for a specific Sturm-Liouville problem \cite{zettl96}, showing that in a one-dimensional system with Dirichlet boundary conditions and no interactions with the wall the energy eigenvalues are a decreasing function from the coordinate of the boundary.

        A simple generalization to this no-go theorem is as follows. A system which has interactions with the walls that satisfy the inequality $U_{\rm ex}(\bm{x_j};\Omega')\leq U_{\rm ex}(\bm{x_j};\Omega)$ cannot have negative pressure.         This is easy to see as it would make the second term of \eqref{paper-eq:press_quant} positive, resulting in an overall positive pressure.

\section{Systems with negative pressure}

    \subsection{Quantum ideal gas with particle-wall interaction and negative pressure}
        Around \eqref{paper-free} we discussed that equilibrium classical states for interacting particles without particle-wall coupling cannot possess negative pressure. The no-go theorem confirmed this conclusion. Let us now turn to scenarios, where the interparticle interaction is absent (i.e. the gas is ideal), but an attractive particle-wall interaction $U_{\rm ex}$ is present; cf. \eqref{paper-eq:hamiltonian}. Such cases are prone to negative pressure, as we show. Consider a single quantum particle in a one dimensional potential well $\Omega\equiv[-a,a]$ where $U_{\rm ex}(x,\Omega) = u(a-|x|)$ \revision{for an arbitrary $u$ with the potential ground set at the center of the well} $u(a)=0$. Using $\pd_x \frac{x}{|x|}=2\delta(x)$, we get from \eqref{paper-eq:press_quant}
        \begin{equation}\label{paper-eq:one_dim_press}
            P = {\sum}_np_n\bigg[\frac{|\psi_n'(0)|^2}{2}+E_n|\psi_n(0)|^2 \bigg].
        \end{equation}
        For the derivation of this result with a more general wall potential in a multidimensional case see S.M. \S 3 \cite{supp}.
        If the particle is in a bound state $\ket{m}$ (due to the potential $U_{\rm ex}$), we take $p_m=1$ and $E_m < 0$ in \eqref{paper-eq:one_dim_press}. Assuming that the state is symmetric $\psi_m'(0)=0$, we find from \eqref{paper-eq:one_dim_press}: $P=-E_m|\psi_m(0)|^2 < 0$. As a concrete example of this negative-pressure effect take
        \begin{equation}
            u(a-|x|)=-u_0\delta(a-b-|x|),\quad 0<b<a, 
        \end{equation}
        where $u_0>0$.
        There is a single symmetric ($\psi_m'(0)=0$) bound ($E_m<0$) state:
        \begin{equation}
            \psi_m(x) = \begin{cases}
                A\sinh{\big[k(a+x)\big]} & -a \leq x \leq b-a, \\
                B\cosh{\big[kx\big]} & b-a \leq x \leq a-b, \\
                A\sinh{\big[k(a-x)\big]} & a-b \leq x \leq a,
            \end{cases}
        \end{equation}
        where $k=\sqrt{2|E_m|}$, $B = A\frac{\sinh{[kb]}}{\cosh{[k(a-b)]}}$ and the boundary conditions at $x=\pm a$ are obeyed. The energy $E_m$ is found from
        \begin{equation}
        \label{paper-bingo}
            \coth{\big[kb\big]}+\tanh{\big[k(a-b)\big]} = {2u_0}/{k}.
        \end{equation}
        Solving \eqref{paper-bingo} numerically we confirm that the pressure is negative (i.e. $\pd_a E_m > 0$) and equals to \eqref{paper-eq:one_dim_press}.
        
    \subsection{Classical ideal gas with negative pressure}
    
        Here is a model of classical ideal gas (no-interparticle interactions) which has negative pressure states. This gas is confined in a disc with radius $a$: $\Omega\equiv\{x_1^2+x_2^2\leq a^2 \}$. Assume the interaction potential between a \revision{$\d l$ part} of the wall and a gas particle at distance $r$ is \revision{$u(r)\d l$, where} $u(r) = -\frac{\sigma}{a} e^{-r^2}$ \revision{and} $\sigma$ is the density of the wall particles for a unit radius. Here for a disc with radius $a$ the density will be $\frac{\sigma}{a}$ to keep the number of particles of the walls $M=\int_0^{2\pi}\frac{\sigma}{a} a\d\varphi$ constant.

        \begin{figure}[ht]
            \centering
            \includegraphics[width=0.35\textwidth]{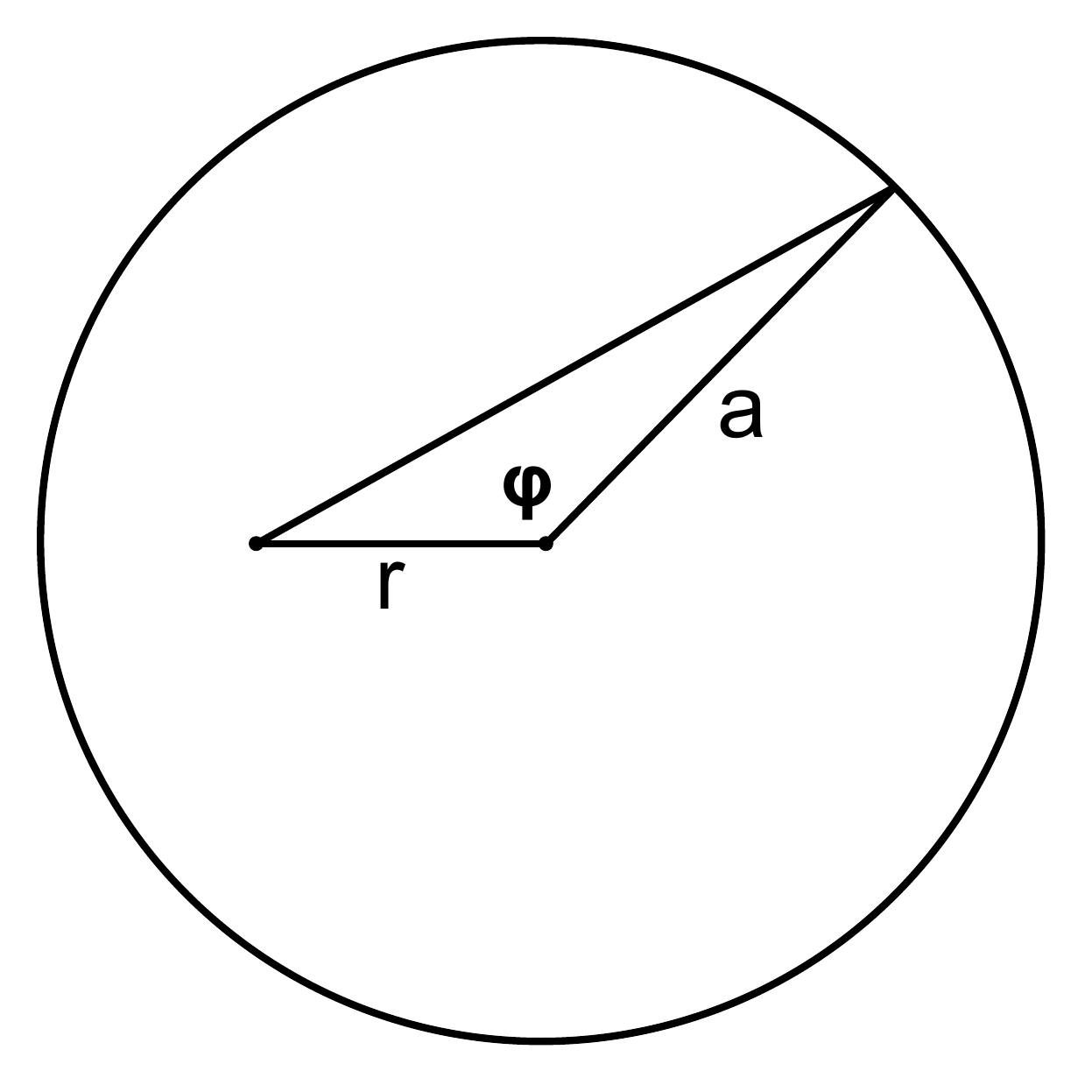}
            \caption{\revision{A schematic diagram of the gas confined in a disc, showing the variables $a$ (radius), $r$ (distance from center) and $\varphi$ used in the integration over the disc walls.}}
            \label{paper-fig:int}
        \end{figure}
        
        The \revision{total} potential of interaction between a gas particle and the \revision{entire} disc walls reads 
        \begin{equation}
        \label{paper-oro}
            U(r) = \int_0^{2\pi} u\bigg(\sqrt{r^2 + a^2 - 2ar\cos{\varphi}}\bigg)a\d \varphi.
        \end{equation}
        Once the gas is ideal, I restrict myself to a single gas particle. Eq. \eqref{paper-oro} produces
        \begin{equation}
            U(r) = -2\pi\sigma e^{-r^2-a^2}I_0(2ar),
        \end{equation}
        where $I_0$ is the zero-order modified Bessel function of the first kind. Coupling this system with a thermal bath to form a canonical ensemble results in [cf. \eqref{paper-free}]
        \begin{equation}
            P =\pd_a \ln \int_0^a re^{-\beta U(r)} \d r.
            \label{paper-krot}
        \end{equation}
        
\begin{figure}[ht]
            \centering
            \includegraphics[width=0.48\textwidth]{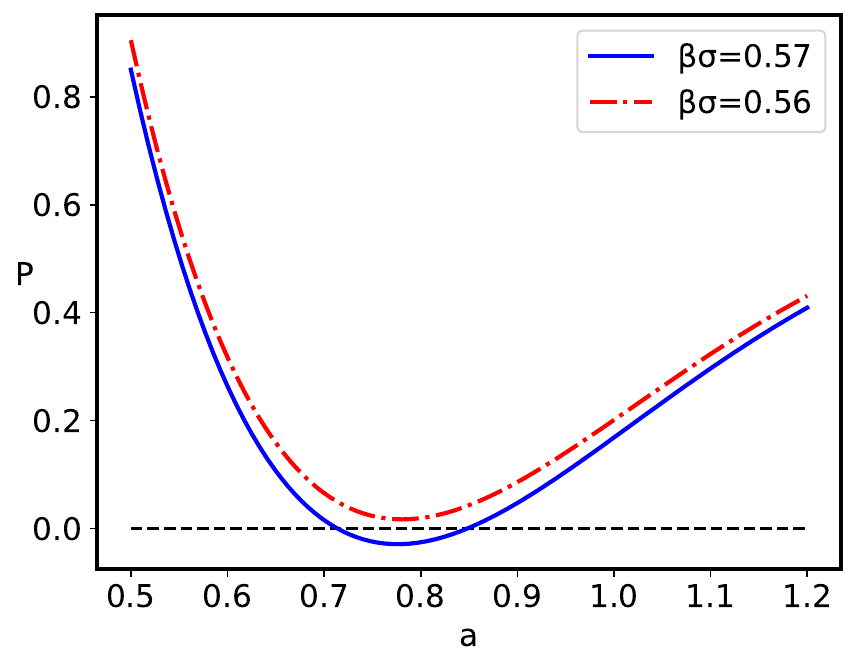}
            \caption{The graph of the pressure depending on the radius of the disc calculated numerically for the values $\beta\sigma=0.57$ (blue, solid) and $\beta\sigma=0.56$ (red, dashed-dotted).}
            \label{paper-fig:1}
        \end{figure}

        The model has two parameters: the system's size $a$ and $\beta\sigma$. Numerical integration shows that \eqref{paper-krot} can turn negative for several values of parameters; e.g. for $a\approx0.78$ there is a transition from positive to negative pressure when changing $\beta\sigma$ from 0.56 to 0.57; see Fig. \ref{paper-fig:1}. Thus we see that (contrary to the common belief) even a classical ideal gas can demonstrate negative pressure states. Eqs. (\ref{paper-oro}--\ref{paper-krot}) can be used in future research as a simple model for negative pressure states.
        
        In the context of the above two examples, I emphasize that the found absolute negative pressure states are stable \revision{(i.e., perform finite motion)} because we assume that other factors keep the walls fixed (or at least slowly moving). Hence the stability scenario is similar to what happens in cases, where the external and internal pressures acting on a wall are different, but the motion of the wall can be neglected (e.g. because it is heavy). 

\section{Results and Discussion}
    
    Negative pressure is an extreme thermodynamic effect with actual and potential applications in natural and artificial technologies. However, misunderstandings about its existence conditions hinder applications. Our results provide insight into existing ideas about negative pressure, as well as intuition and methodology for future studies. The no-go theorem result prohibits negative pressures in any non-equilibrium state and for any Hamiltonian assuming no particle-wall interaction. When the no-go condition is avoided there are systems that show negative pressures. Two examples, quantum and classical, provide mechanisms to achieve negative pressure.  
    
    Additionally, the no-go theorem sheds light on systems that are often seen as demonstrating negative pressures. One such system is the van der Waals liquid. This known phenomenological model supposedly demonstrates negative pressure in the phase diagram region prohibited by the thermodynamic stability (i.e. by Maxwell's construction) \cite{landau5}. The accepted interpretation of states from this region is that they are metastable (non-equilibrium) states \cite{lebowitz71,larralde06}. Since the van der Waals model does not account for the particle-wall interaction, I conclude that the metastable-state negative pressure in the model contradicts the no-go theorem. There are several ways of resolving this contradiction. First of all, this can be due to the limitations of the no-go theorem itself; see {\bf 2} and {\bf 3} above. Such scenarios do not seem likely to me. \revision{Limitations related to {\bf 3} are discussed after \eqref{paper-ado}. As for possible caveats in {\bf 2}, recall that the equilibrium definition of pressure does not depend on the process. Hence, the way we change the boundary should not affect the pressure. Note in this context that statistical mechanical descriptions of metastability involve discrete-state Markov processes \cite{metastability_schulman,larralde06}, and it is known that the theory of open quantum systems describes such systems \cite{petr}. This supports the hypothesis that the physics behind \eqref{paper-dag} can also account for classical metastable systems.  
    }
    
    A more likely scenario is that the negative pressure is an artifact of approximations involved in formulating the van der Waals model and its metastable states \cite{landau5,bocquet}. Metastable states with negative pressures have been observed experimentally \cite{caupin13, balibar00, imre98} and are important in nature \cite{brown13, miller68}. To our understanding, such systems involve particle-wall interactions, e.g. such interactions are certainly present for a self-consistent wall (liquid bubble). Indeed, molecular dynamics shows that model liquids can have negative pressure bubbles \cite{matsumoto98}. In the light of my results, one scenario for such bubbles is that the Lennard-Jones potential employed in molecular dynamics leads to interaction between bubble particles and their effective boundary. This would be sufficient to avoid the no-go condition as well as it would provide a mechanism for nonequilibrium negative pressure states in liquids. Finally, it is important to note that there are reports on negative pressure in equilibrium systems that combine mean-field methods with particle-wall interactions \cite{podgornik91, podgornik24}. These refer to a polyelectrolyte inside a charged confinement, where there are conformational changes accompanied by changes in the pressure sign \cite{podgornik91, podgornik24}. Such results do not contradict the no-go theorem but still invite further considerations. \revision{ Additional considerations are also needed for the statistical mechanics of self-gravitating point particles, where non-stationary regimes of negative pressure can be possible. Such systems can be unstable (i.e., their space-density is not integrable at small distances) due to the singular, unscreened character of the Newtonian gravitation at small distances \cite{padmanabhan1990}.}

\section*{Acknowledgements}
    I am thankful to V. Abgaryan, A.E. Allahverdyan, K. Hovhannisyan, Y.Sh. Mamasakhlisov and  V.E. Mkrtchian for thoughtful discussions and constant support. This work was supported by the Higher Education and Science Committee, Republic of Armenia grants No. 22AA-1C023, 21AG-1C038 and 24FP-1F030.

%\clearpage
\bibliography{paper}
    
\end{document}